\documentclass[a4paper,11pt]{article}
\pdfoutput=1
\usepackage[utf8x]{inputenc}
\usepackage{jheppub} 
\usepackage{soul}
\usepackage{amsmath,amssymb}
\usepackage{mathrsfs}
\usepackage{subfig}
\usepackage{float}
\usepackage{slashed}
\usepackage{multirow}
\usepackage{ulem}
\usepackage{soul}

\title{Charged Higgs effects in IceCube: PeV events and NSIs}

\author[a]{Ujjal Kumar Dey,}
\author[b]{Newton Nath,}
\author[c,d]{Soumya Sadhukhan}

\affiliation[a]{Department of Physical Sciences, Indian Institute of Science Education and Research Berhampur,\\ Transit Campus, Government ITI, Berhampur 760010, Odisha, India}

\affiliation[b]{Instituto de F\'{i}sica, Universidad Nacional Aut\'{o}noma de M\'{e}xico,\\ A.P. 20-364, Ciudad de M\'{e}xico 01000, M\'{e}xico}

\affiliation[c]{Ramakrishna Mission Residential College (Autonomous),\\ Narendrapur, Kolkata 700103, India}

\affiliation[d]{Department of Physics and Astrophysics, University of Delhi,\\ Delhi 110 007, India}

\emailAdd{ujjal@iiserbpr.ac.in}
\emailAdd{newton@fisica.unam.mx}
\emailAdd{physicsoumya@gmail.com}

\abstract{
Extensions of the Standard Model with charged Higgs, having a non-negligible coupling with neutrinos, can have interesting implications vis-\`{a}-vis neutrino experiments. Such models can leave their footprints not only in the ultra-high energy neutrino detectors like IceCube but can also give rise to the neutrino non-standard interactions (NSIs). We consider a model based on the neutrinophilic two-Higgs doublets and study its imprints in the excess neutrino events in the 1-3 PeV energy bins at the IceCube. This is facilitated by the existence of a charged scalar in the model which can result in a Glashow-like resonance. The same charged scalar can be responsible for sizeable NSIs. We perform a combined study of the latest IceCube data along with various other constraints arising from different neutrino experiments together with the limits set by the LEP experiment, and explore the parameter space which can lead to a sizeable NSI. 
}

\allowdisplaybreaks

\begin{document}
\maketitle
\flushbottom

\section{Introduction}
\label{sec:intro}
The perplexing nature of neutrinos that can not be accommodated within the Standard Model (SM) of particle physics has been giving the impetus to research in particle physics for quite some time. After the discovery of neutrino oscillations~\cite{Zyla:2020zbs} which established the massive nature of neutrinos and their mixing parameters, a host of experiments are dedicated to measure neutrino oscillation parameters with increasing precisions. In addition to these, currently operating ultra-high energy (UHE) neutrino detector at the south pole, i.e., the IceCube neutrino observatory has detected highest energy neutrinos till date~\cite{Aartsen:2013bka, IceCube:2021rpz} and thus ushered a new era in neutrino physics.
The IceCube collaboration~\cite{Aartsen:2014gkd, Schneider:2019ayi} has reported, the  presence of three neutrino events with energies 1-3 PeV, and very recently one more event around the long sought-after Glashow resonance energy $\sim$6 PeV~\cite{IceCube:2021rpz}.
The detection of the neutrinos with such a high energy inevitably bears our high hopes of searching  physics beyond SM (BSM), the onus of which, till now, was mostly on high energy collider experiments. Not only a number of recent studies considered the opportunities of utilising IceCube in exploring various BSM scenarios like, leptoquarks~\cite{Dey:2015eaa, Chauhan:2017ndd, Becirevic:2018uab}, supersymmetry~\cite{Dev:2016uxj}, dark matter~\cite{Rott:2014kfa, Esmaili:2014rma, Allahverdi:2015ssa, Bhattacharya:2016tma, Dev:2016qbd, Cohen:2016uyg, Chianese:2017nwe, Borah:2017xgm, Kachelriess:2018rty, Sadhukhan:2018nsk, Sui:2018bbh, Chianese:2018ijk, Chauhan:2018dkd, Guo:2020drq}, probing the possibility of neutrino decay~\cite{Denton:2018aml, Abdullahi:2020rge} etc., but studies of combining the results of LHC and IceCube have also been made~\cite{Dutta:2015dka, Dey:2017ede, Pandey:2019apj}.   
Attempts of solving the enigma of neutrino masses generated, and are still generating, a formidable literature of phenomenological models (see~\cite{King:2015aea, Cai:2017jrq} for a review). Many of these scenarios can induce neutrino non-standard interactions (NSI)~\cite{Wolfenstein:1977ue}, a possible sub-leading effect that may affect the propagation of neutrinos, in presence of some heavier messenger fields. From the low energy point of view, when these fields are integrated out NSIs can be generated in the form of dimension-6~\cite{Buchmuller:1985jz, Bergmann:1998ft, Bergmann:1999pk, Wise:2014oea} and/or -8~\cite{Berezhiani:2001rs, Davidson:2003ha} effective operators (for further details see~\cite{Ohlsson:2012kf, Miranda:2015dra, Valle:1987gv, Roulet:1991sm, Guzzo:1991hi}). Moreover, there are studies which explore the possibility of probing NSIs at collider experiments~\cite{Davidson:2011kr, Choudhury:2018xsm, Babu:2020nna}.
There is a class of extended scalar sector models with neutrinophilic couplings that strive to address the problem of neutrino mass~\cite{Ma:2000cc, Gabriel:2006ns, Davidson:2009ha, Machado:2015sha, Babu:2019mfe}. 
A concise study of NSIs, from the point of view of both the effective dimension-6 operators and their realizations in terms of simple renormalizable models involving leptoquarks, bileptons, as well as leptophilic doublet scalar, is performed in~\cite{Wise:2014oea}.
Here we focus on a variant of neutrinophilic two Higgs doublet models ($\nu$2HDM), as we have proposed in our previous work~\cite{Dey:2018yht}, that can lead to sizeable NSIs owing to the presence of couplings between neutrinos and the charged Higgs present in the model.
In this work, we also make an attempt to give an explanation of the observed neutrino events in the 1-3 PeV energy range reported by the IceCube collaboration~\cite{Aartsen:2014gkd, Schneider:2019ayi}. Note that some of the recent studies incorporating NSIs in view of IceCube data are performed in~\cite{Esmaili:2013fva, Salvado:2016uqu, Aartsen:2017xtt, Demidov:2019okm}. 
Here, the standard $\nu$2HDM has been modified in a way such that the second scalar doublet $\Phi_{2}$ couples only to the electrons and neutrinos. We achieve this by assigning a negative charge to the $e_{R}$ under a global $U(1)$ symmetry. The physical charged Higgs, arising due to the second doublet, will have coupling with electron and neutrino. This can give rise to the interesting possibility of producing the charged Higgs resonantly from the interaction of the incoming ultra-high energy neutrino and the electron. This Glashow-like resonance can give rise to specific signature at the IceCube.
On the other hand the coupling of doublet $\Phi_{2}$ with the left-handed lepton doublet and the right-handed electron leads to dimension-6 four fermion operators involving neutrino of the form, 
\begin{equation}
\label{eq:NSI}
\mathcal
{L}_\text{NSI} \supset 
(\overline{\nu}_\alpha \gamma^{\rho} P_{L}\nu_\beta)
(\bar{e} \gamma_{\rho} P_{R}e)\epsilon_{\alpha\beta} 
+ \text{H.c.} \;,
\end{equation}
where $\epsilon_{\alpha\beta}$ represent NSI parameters with $\alpha, \beta = e, \mu, \tau$, once the heavy charged Higgs has been integrated out from the theory. Interestingly, we notice that this framework only leads to two sizeable NSIs $ \epsilon_{ee}, $ and $ \epsilon_{e\tau} $. 
Note that in a similar doublet extension model all the diagonal NSI parameters are explored where one of them was found to be as large as one percent level~\cite{Wise:2014oea}.
It goes without saying that any BSM scenario undergoes various phenomenological constraints coming from numerous particle physics experiments. Using various limits, we first investigate the range of charged Higgs mass that can explain the IceCube PeV neutrino events. 
Furthermore, considering the combined effect of different experimental constraints coming from the IceCube, Oscillation+COHERENT~\cite{Coloma:2017ncl}, Borexino~\cite{Agarwalla:2019smc}, TEXONO~\cite{Deniz:2010mp} as well as from the numerical results of DUNE+T2HK~\cite{Acciarri:2015uup,Abe:2018uyc}, we examine the allowed parameter space that can lead to possible NSIs.
In addition, we also maintain the limits arising from the LEP and other observations~\cite{Abbiendi:2013hk, LEP:2003aa, Fox:2011fx}. Most of these constraints are already presented in our previous work~\cite{Dey:2018yht}. The interesting thing about this study is the effect of the charged Higgs presenting a synergy between the UHE neutrino events observed at the IceCube and the low energy neutrino oscillation experiments. 
While in this paper we stick to a basic model set-up, it can be viewed as a first-step towards the generalisation of multiple charged Higgs scenarios (e.g., the two charged Higgs scenario of Zee model~\cite{Babu:2019vff}), where basic tenets of charged Higgs and its appreciable coupling with neutrinos are present. 
The remainder of the paper is organized as follows. In the next Sec.~\ref{sec:model}, we provide a brief description of our model that leads to neutrino non-standard interactions. Different prospects of IceCube have been discussed in Sec.~\ref{sec:ic}. Firstly in Sec.~\ref{sec:PeV-ic}, it has been shown that how one can explain observed PeV neutrino events in this model. In Sec.~\ref{sec:nsi-IC}, the sensitivity of IceCube to test NSIs has been pointed out by incorporating various other experimental limits. Finally, we summarize our findings in Sec.~\ref{sec:concl}.

\section{Model Description and NSIs}
\label{sec:model}
Neutrinophilic 2HDMs, be it $\mathbb{Z}_{2}$-symmetric with possible Majorana masses~\cite{Ma:2000cc, Gabriel:2006ns} or $U(1)$-symmetric with exclusive Dirac neutrino masses~\cite{Davidson:2009ha}, do not give rise to NSIs. This is simply because of the fact that two different Higgs doublets provide masses to the neutrinos and remaining massive SM particles, rendering the non-existence of interaction between the left-handed neutrinos and charged leptons as well as quarks. Elaborate scalar extensions larger than the two-doublet paradigm where large NSIs can arise have been studied in~\cite{Forero:2016ghr}.
In~\cite{Dey:2018yht} we have shown that even within the two-doublet scenario one can have sizeable NSIs. This can be achieved by imposing only the right-handed electron $e_{R}$ a global $U(1)$ charge of ($-1$) which basically boils down to giving mass to the electron along with the neutrinos through the second doublet $\Phi_{2}$. 
We list the relevant  $SU(2)_L$ and global $U(1)$ charges  for the necessary fields in Table~\ref{tab:charges}. 
%
\begin{table}[h]
    \centering
    \begin{tabular}{|c|c|c|c|c|c|c|c|c|c|c|c|}\hline
   &$L_e$ &$L_\mu$ &$L_\tau$ & $e_R$ & $\mu_R$   &$\tau_R$  & $\nu_{eR}$ & $\nu_{\mu R}$ & $\nu_{\tau R}$ & $\Phi_1$ &$\Phi_2$  \\ \hline \hline 
   $SU(2)_L$ & 2 & 2 & 2 & 1 & 1 & 1 & 1 & 1 & 1 &  2 & 2 \\ \hline
   $U(1)$ & 0 & 0 & 0 & $-1$ & 0 & 0 & 1 & 1 & 1 & 0 & 1  \\ \hline
    \end{tabular}
    \caption{\footnotesize The $SU(2)_L$ and global $U(1)$ quantum numbers of the leptons and Higgs doublets in the model.}
    \label{tab:charges}
\end{table}
%
The generic Yukawa coupling relevant for our discussion can be written as,
\begin{align}
\label{eq:mdlLag}
\mathcal{L}_{\nu\text{2HDM}} \supset 
		y_{\ell_{i}e}\bar{L}_{\ell_{i}}\Phi_{2}e_{R} +
		y_{\ell_{i} \nu}\bar{L}_{\ell_{i}}\tilde{\Phi}_{2}\nu_{\ell_{i} R} + \text{H.c.} \;,
\end{align}
where $L_{\ell_{i}} = (\nu_{\ell_{i}}~~\ell_{i})^{T}$ with $(\{\ell_{1},\ell_{2},\ell_{3}\} = \{e,\mu,\tau\})$ represents the SM lepton doublets. The first term of Eq.~\eqref{eq:mdlLag} gives the charged Higgs interactions $y_{\ell_{i}e} \bar{\nu}_{\ell_{i} L}H^{+}e_{R}$ which result in a $t$-channel $H^{+}$ mediated process. After integrating out the mediator and using appropriate Fierz transformation we obtain the neutral-current NSI terms between the SM neutrinos and electrons. Note that as compared to our previous work, we make a slight change of notation, the Yukawa couplings here $\{y_{ee}, y_{\mu e}, y_{\tau e}\}$ map to $\{y_{e}, y_{1}, y_{2}\}$ in~\cite{Dey:2018yht}. The general form of this NSI parameters can be written as,
\begin{align}
\label{eq:gennsimodel}
\epsilon_{\ell_{i}\ell_{j}} = \frac{1}{2\sqrt{2}G_{F}}
		\frac{y_{\ell_{i}e}y_{\ell_{j}e}}
		{4m^{2}_{H^{\pm}}} \;,
\end{align}  
where $G_{F}$ is the Fermi constant. 
Following the definition of the neutral-current NSI~\cite{Wolfenstein:1977ue}, the respective dimension-6 four-fermion operator in our model can be written as,
%
\begin{align}
\label{eq:ncnsi}
\mathcal{L}_{\rm NSI}^{\rm NC} = -2\sqrt{2}G_{F}
		(\bar{\nu}_{\alpha}\gamma^{\rho}P_{L}\nu_{\beta})(\bar{e}\gamma_{\rho}P_{R}e)\epsilon_{\alpha \beta} + {\rm H.c.},
\end{align}
where $\alpha, \beta = e, \mu, \tau$. 
Here, the NSI matrix $ \epsilon_{\alpha\beta} $ can be written as
\begin{align}
\label{eq:matternsi}
\epsilon_{\alpha\beta}\ =  \left(\begin{array}{ccc}
 \epsilon_{ee} & \epsilon_{e\mu} & \epsilon_{e\tau} 
\\
\epsilon^{*}_{e\mu}  & \epsilon_{\mu\mu} & \epsilon_{\mu\tau}
\\
\epsilon^{*}_{e\tau} & \epsilon^{*}_{\mu\tau} & \epsilon_{\tau\tau}
\end{array}\right)\, ,
\end{align}
where $ \epsilon_{\alpha\beta} = |\epsilon_{\alpha\beta}| e^{i\phi_{\alpha\beta}} $ for $ \alpha \neq\beta $. 
%
%
Coming back to the general NSI parameter $\epsilon_{\ell_{i} \ell_{j}}$ in our model, note that the complex nature of the Yukawa couplings $y_{\ell_{i}e}$ can, in principle, generate phases for the off-diagonal NSI parameters. The phenomenological constraints coming from the LEP experiment, lepton flavour violation processes, magnetic moments, BBN etc., on various parameters e.g., the vacuum expectation values (vev) $v_{1,2}$ of the two doublets $\Phi_{1,2}$, the Yukawa couplings and their combinations are detailed in our previous study~\citep{Dey:2018yht}. It is worth mentioning here that in general the NSI matrix, as given by Eq.~\eqref{eq:matternsi}, consists of three real diagonal and three off-diagonal complex entries, however within our framework, after considering various phenomenological constraints, there are only two sizeable NSI parameters, namely $\epsilon_{ee}$ and $\epsilon_{e\tau}$. 
As we have discussed previously in~\cite{Dey:2018yht}, depending on various allowed Yukawa couplings one can in principle have all six NSI parameters. Each of these parameters actually depends on the concerned Yukawa coupling. The constraints coming from $\mu \to 3 e$ decay puts stringent bound on the Yukawa $y_{\mu e} \sim 10^{-6}$. This makes all the NSI parameters $\epsilon_{x\mu}~(x = e,\mu,\tau)$ negligible as is evident from Eq.~\eqref{eq:gennsimodel}. Therefore, only possible sizeable NSIs in this model are $\epsilon_{ee}, \epsilon_{e\tau}$ and $\epsilon_{\tau \tau}$. Since the available bounds on $\epsilon_{\tau \tau}$ are anyway too relaxed~\cite{Ohlsson:2012kf}, we are not considering it. Other lepton flavour violating (LFV) processes e.g., $\ell_{\alpha} \to \ell_{\beta}\gamma$ lead to relatively less stringent constraints on the relevant Yukawas, making $\epsilon_{ee}$ and $\epsilon_{e\tau}$ the only sizeable NSIs in this model. These LFV constraints are considered in detail in the previous work~\cite{Dey:2018yht}. In the current work we maintain all those constraints and the benchmark points, if and when needed, respect them as well. After this brief discussion of our model set-up and the relevant notations and conventions, we shall now discuss the signatures and prospects of probing the charged Higgs at the IceCube by focusing on the PeV events and  NSIs.   

%

\section{Signatures and Prospects at IceCube}\label{sec:ic}
The IceCube detector is one of the major neutrino telescopes that can detect ultra-high energetic (UHE) neutrino events, preferably coming from astrophysical sources like the supernovae, gamma ray bursts (GRB), active galactic nuclei (AGN) or from other possible new cosmic sources~\cite{Aartsen:2015rwa, Aartsen:2017mau}. 
The IceCube experiment, after a run of 7.5 years, has recorded a total of 103 neutrino events, combining both the track and shower events with neutrino interaction vertex inside the detector, of which  60 events are above 60 TeV~\cite{deWasseige:2019pez}.
Among high energy starting events (HESE) at IceCube, track events arise from the charged current $\nu_{\mu}$, whereas charged currents of $\nu_e$ and $\nu_{\tau}$ and neutral currents of all neutrino flavours result in shower events~\cite{Aartsen:2015rwa}. Moreover, the IceCube has already detected several higher energy neutrino events till date in the PeV energy range~\cite{Aartsen:2014gkd, Schneider:2019ayi, IceCube:2021rpz}. In the following we would strive to explain some of these events in our scenario.

\begin{figure}[!htbp]
\hspace{2cm} \includegraphics[scale=0.55]{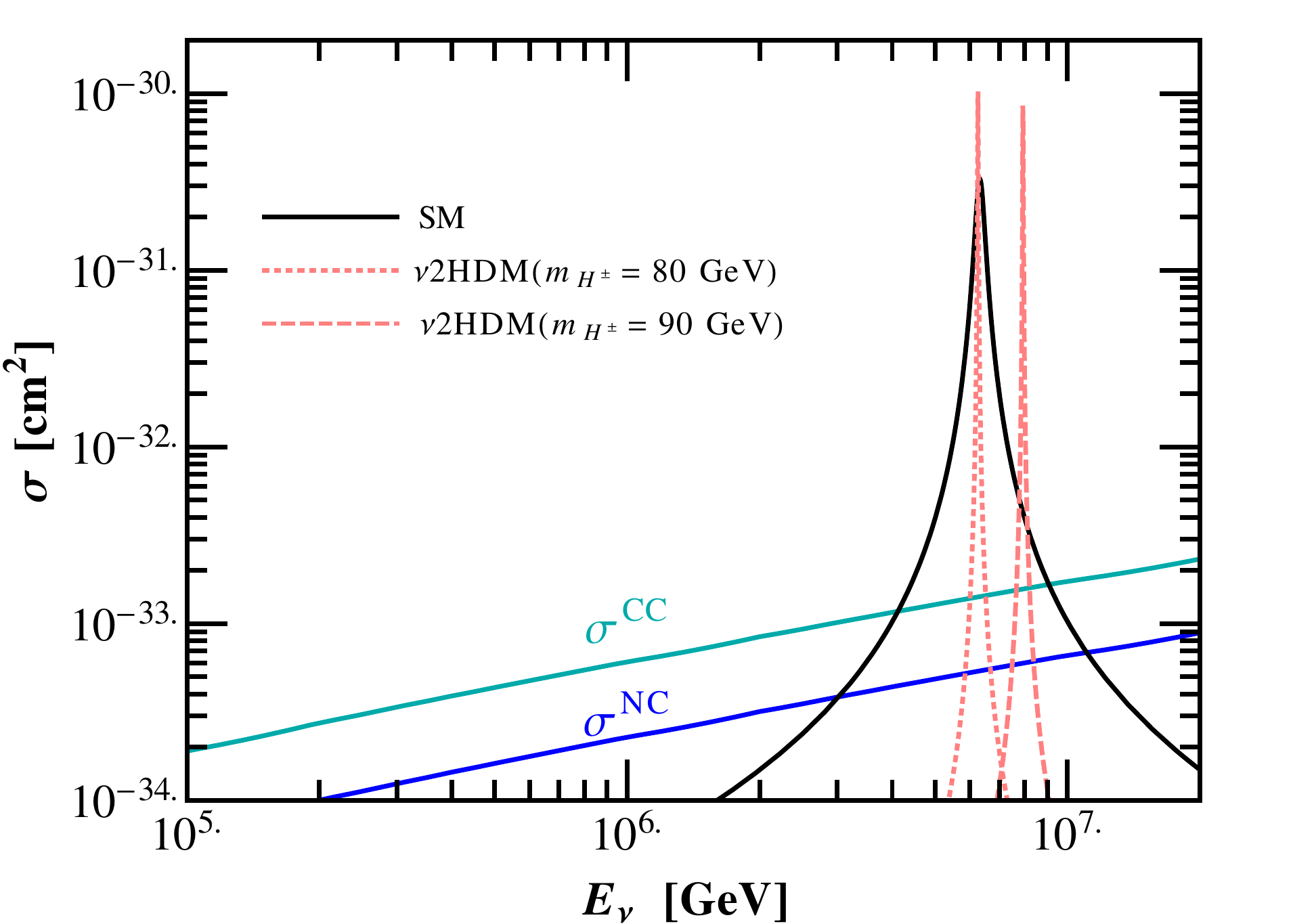} 
\caption{\footnotesize Neutrino scattering cross sections for different cases. The SM deep inelastic neutrino-nucleon cross sections are presented by the blue (cyan) curve for neutral (charged) current interaction. The black solid curve shows the Glashow resonance. The pink dotted and dashed curves represent the resonance case originated due to the charged Higgs  corresponding to two benchmark masses of 80 and 90 GeV, respectively.}
\label{fig:CrossSection}
\end{figure}
%

\subsection{PeV Neutrino Events at IceCube}
\label{sec:PeV-ic}
The basic premise of finding a signature of charged Higgs at IceCube lies in the neutrino-electron coupling, mediated by the charged Higgs. This coupling is facilitated by the first term in Eq.~\eqref{eq:mdlLag}. Incoming neutrinos with sufficiently high energy can interact with electrons and resonantly produce the charged Higgs, $H^{-}$. Clearly, this resonant charged Higgs production due to the neutrino-electron interaction is reminiscent of the renowned Glashow resonance~\cite{Glashow:1960zz}. 
%

%
The effective neutrino-electron cross section in the presence of a resonant charged Higgs can be written as
\begin{equation}
\sigma_{H^{\pm}}(E_{\nu}) = \frac{8 \pi}{m_{H^{\pm}}^2} \left[\frac{2 m_e E_{\nu}\text{Br}(H^{\pm} \to e \nu_e) \text{Br}(H^{\pm} \to \text{all}) \Gamma_{H^{\pm}}^2 }{ \left[(2 m_e E_{\nu} - m_{H^{\pm}}^2)^2 + \Gamma_{H^{\pm}}^2 m_{H^{\pm}}^2 \right]}\right],
\end{equation} 
where the total charged Higgs decay width in our case is given by
\begin{equation}
\Gamma_{H^{\pm}} = y_{ee}^2 \frac{m_{H^{\pm}}}{16 \pi} \;.
\end{equation}
In Fig.~\ref{fig:CrossSection} we show the typical cross section of neutrino scattering with nucleons and electrons. The usual neutrino-nucleon deep inelastic scattering (DIS) cross sections due to the neutral current (NC) and charged current (CC) interactions are shown by the blue and cyan solid curves, respectively
 \footnote{
In order to demonstrate the cross sections for DIS processes, we use \texttt{nCTEQ} parton distribution functions~\cite{ncteq} and references therein.}.
The black solid curve depicts the Glashow resonance originated from the $W$ mediated neutrino-electron scattering. In the similar vein, the charged Higgs mediated neutrino-electron scattering, typical to our model set-up, results in Glashow-like resonance structures in the neutrino-electron scattering cross section. Note that there will be no interference between the current process and the standard Glashow resonance owing to the fact that in the present scenario only the right-handed electrons are involved. We show them by the pink dotted and dashed curves corresponding to two benchmark charged Higgs masses of 80 and 90 GeV respectively, where the relevant Yukawa coupling $y_{ee} = 0.28$ is taken. 
It is obvious that the charged Higgs decay widths manifest themselves through the widths of the resonant peaks which dominantly depend on the relevant Yukawa coupling and mass. 
In the following we will utilise these charged Higgs resonances to have a better fit for the IceCube event distribution as compared to the SM interpretations.

For the numerical analysis of the observed event distribution at IceCube, next we need the information of the incoming neutrino flux. Apart from the atmospheric components, we take the astrophysical neutrino flux with the single power law behaviour in energy, to analyse the observed diffuse neutrino flux. In an isotropic astrophysical neutrino flux, the initial flavour components can be different for different astrophysical sources. Assuming pion decay production mechanism for the neutrinos, the predicted flavour ratios in the initial flux are expected to be $\Phi_{\nu_{e}}:\Phi_{\nu_{\mu}}:\Phi_{\nu_{\tau}} =  1:1:1$ at the detector~\cite{Stachurska:2019srh, Beacom:2003nh}. The initial astrophysical neutrino flux for all flavours, with single component is parametrized as
\begin{equation}
\Phi (E_{\nu}) = \Phi_{\text{ast}}^{0} \left(\frac{E_{\nu}}{100 \ \text{TeV}} \right)^{-\gamma},
\label{inif}
\end{equation}
where $\Phi_{\text{ast}}^{0}$ is the flux amplitude and $\gamma$ is the spectral index, both of which are determined by astrophysical considerations. The current IceCube data on observed events best-fit this flux when these two parameters are taken as $\Phi_{\text{ast}}^{0} = 6.45 \times 10^{-18} (\rm GeV \ cm^2 \ s \ sr )^{-1}$ and $\gamma = 2.89$, as shown in Ref.~\cite{Schneider:2019ayi}.  
There can be two major inconsistencies in the IceCube best-fit spectrum~\cite{Schneider:2019ayi}. 
%
Firstly, with the SM particle content we expected to see a number of Glashow resonance events. Those were predicted as the resonant peak due to a $W^{-}$ production in the process ${\bar \nu}_e e^- \to W^-$. There are some explanations of the absence of these Glashow events up to now, as explained in the Refs.~\cite{Sadhukhan:2018nsk, Mohanty:2018cmq, Huang:2019hgs}. Along with the recent observation of one Glashow resonance event at the IceCube~\cite{IceCube:2021rpz}, small amount of data and the associated experimental error in the current run render it insignificant enough to be addressed right now.
Secondly, 7.5 years of IceCube run has indeed observed a number of events at energy $E_{\nu} \gtrsim 1$ PeV, in the initial super-PeV bins~\cite{Schneider:2019ayi, Aartsen:2014gkd}. The best-fit IceCube parameters do not explain the presence of these events in the spectrum. Statistically, they are a good fit to have an SM prediction within the errors of the observation, but do not address readily the presence of real observed events. This is one of the major point of contention that compels us to look for BSM explanation of these events. The presence of these isolated events direct us towards exploring new resonances. 
In the neutrinophilic 2HDM ($\nu$2HDM), a light charged Higgs can be present around 100~GeV region, which can induce a resonance effect similar to the Glashow resonance in the super-PeV energy. We quantify the amount of modification to IceCube events in the $i$-th bin, in presence of this $H^{\pm}$, as 
\begin{equation}
\label{eq:ICevtNum}
N_i = \Omega ~T_0 ~n_{\rm int} \int^{E_{i+1}}_{E_i} dE \left[ \int^{\infty}_E  \Phi (E_{\nu})  {\sigma}_{H^{\pm}} (E_{\nu}) \delta (E - \epsilon E_{\nu}) dE_{\nu} \right], 
\end{equation} 
where $E_{\nu}$ and $E$ represent the incident neutrino energy and the energy at the detector, respectively. Here, $T_0$ is the total time of exposure for the IceCube experiment i.e. the time period over which the experiment is running, 2635 days, where $n_{\rm int}$ represents the total effective number of interaction points inside the IceCube detector. The solid angle coverage $\Omega = 4 \pi$ and $\Omega = 2 \pi$ are used for the sub-PeV and super-PeV IceCube event computation, respectively~\cite{Gandhi:1995tf}. 
Also, $\epsilon$ is the acceptance of the final states in the IceCube detector. The value of the $\epsilon$ in our model, dealing with a charged Higgs decay is taken in the range of $0.20-0.25$, which is the case for leptonic final states. 
Here, the exact relation between the energy of the outgoing electron and that of the incoming neutrino is approximated as a delta function. The delta function $\delta(E - \epsilon E_{\nu})$ effectively ensures that the deposited energy $E = \epsilon E_{\nu}$, where $\epsilon$ is the acceptance which we have varied here in the $0.20-0.25$ range. 
For delta function formalism and acceptance of $e\nu$ final state, we follow Refs.~\cite{Kistler:2013my, Dutta:2015dka}. 
%

\begin{figure}[h]
\hspace{2cm} 
\includegraphics[scale=0.4]{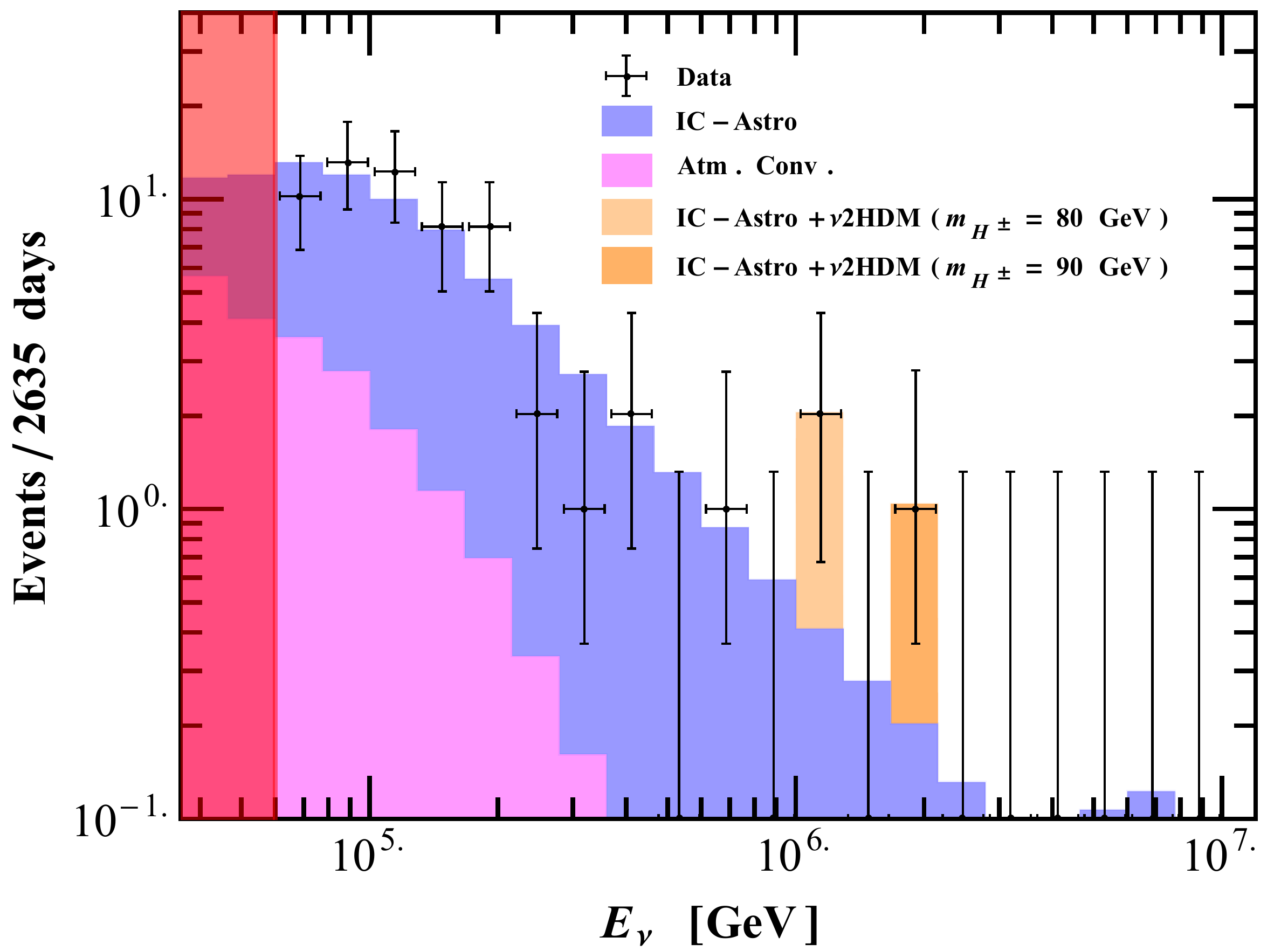} 
\caption{\footnotesize Number of events in each IceCube bins for total run of 2635 days. Event computation for astrophysical best-fit parameters $\Phi_{\text{ast}}^{0} = 6.45 \times 10^{-18} (\rm GeV \ cm^2 \ s \ sr )^{-1}$ and $\gamma = 2.89$. The SM events resulting from astrophysical and atmospheric flux are taken from the IceCube collaboration~\cite{Schneider:2019ayi}, as shown using the the blue, and pink histograms, respectively. Here, the light-orange and dark-orange bars represent the number of events corresponding to 80 and 90 GeV charged Higgs masses, respectively. The latest IceCube HESE data sets are shown using the black cross marks. The vertical red-band depicts the data set which are not included by the IceCube analysis.
}
\label{fig:Events}
\end{figure}

%
In the $\nu$2HDM case, only allowed decay mode is the leptonic $H^{\pm} \to l \nu$ one. Since the leptonic final states are accompanied by neutrinos, visible energy transfer remains low and therefore results in the low acceptance. This is contrary to the case of Glashow resonance where the $W$-boson decays dominantly to the hadronic channels that transfer most of their energy to have a high acceptance. This constitutes another distinguishing feature of our case in comparison to the standard Glashow resonance. Actually, the charged Higgs of our model is electrophilic and has a large ($\simeq$1) leptonic branching ratio whereas the electronic decay branching ratio for the $W$-boson is only around 10\%. Therefore, the $W$-boson  contribution  to the leptonic final states  will be minimal compared to the charged Higgs resonance process.
Using the construction above, the bin-wise event modifications are calculated and effects of the charged Higgs resonances are reflected in Fig.~\ref{fig:Events}. For the two benchmark charged Higgs masses of 80 and 90 GeV, we can explain the events in the first and third super-PeV IceCube bins, which we show using the light-orange and dark-orange bars, respectively.
Evidently we can not fit both the bins simultaneously, since we have only one charged Higgs in the model. The benchmark value for $y_{ee}$ to compute the charged Higgs contribution is taken to be 0.28, which satisfies all current experimental bounds discussed in Ref.~\cite{Dey:2018yht}.
The black crosses represent the latest IceCube HESE data, whereas  the astrophysical, and conventional atmospheric flux components are shown using the blue, and pink histograms respectively, as has been adopted  from \cite{Schneider:2019ayi}. Notice that the data set below 60 TeV, as shown by the vertical red-band, are not included in the IceCube analysis~\cite{Schneider:2019ayi}, and hence we also do not include these in our simulation.
For the 80 GeV mass of $H^{\pm}$, the cross section peak at around $E_{\nu} \sim 5~$PeV which translates (for an acceptance $\sim 0.20$) to two events observed in the first super-PeV bin. Similarly, if we take the charged Higgs mass to be 90 GeV the cross section peak translates to a single event observed by IceCube in the third super-PeV bin. 
The chosen benchmarks for $m_{H^{\pm}}$, apparently coincident with massive SM gauge bosons, do not interfere with the Glashow resonance. As it is mentioned earlier, the Glashow resonance peak and e.g., 80 GeV charged Higgs peak can be at the same incident neutrino energy in the resonant process, their final decay states compositions are quite different.
The leptophilic charged Higgs mediated process we discuss here receives minimal contribution from the $W$-boson mediated resonant decay.
The $m_{H^{\pm}}$ = 80 GeV cannot contribute to the dominant hadronic decay of the Glashow resonance.  
As a possible distinction of the Glashow resonance and the charged Higgs resonance, recall that in the Glashow resonance the final events are predominantly jets as they are coming from a $W^-$ decay. The jets result in fully hadronic showers and their equivalent electromagnetic energy deposit is a large part (90\%) of the incident neutrino energy. Similarly, for the charged current interaction of $ \nu_e $ that produces electrons along with the recoiling nucleon, most of the energy (95\%) gets deposited in the form of either electromagnetic energy or hadronic showers. In both of these cases high energy deposited shower or cascade events are expected. 
On the other hand, for the charged Higgs resonance in the neutrinophilic 2HDM, the final state is electron accompanied with  neutrinos that carry away a large section of energy as missing energy. Therefore, energy of the final state electron is relatively lesser to an extent of 20-25\% of incident neutrino energy (see~\cite{Kistler:2013my}). Even if electron deposits its entire energy in an electromagnetic shower, due to small energy transfer the shower event will be less energetic ones with smaller probability of a cascade type event.
Having said this, if IceCube can not distinguish hadronic shower from electronic shower, even then the Glashow resonance coming solely from the SM has the peak location around 6 PeV, a preliminary observational confirmation of which has recently appeared~\cite{IceCube:2021rpz}. But, in our case the peak location depends on the specific mass of the charged Higgs which we take as a parameter. Also, even if the charged Higgs mass, and corresponding couplings are such that the peak location shifts towards Glashow resonance, the effective number of events will be small enough to overwhelm the SM Glashow peak.

\begin{figure}[t]
\centering
\includegraphics[scale=0.52]{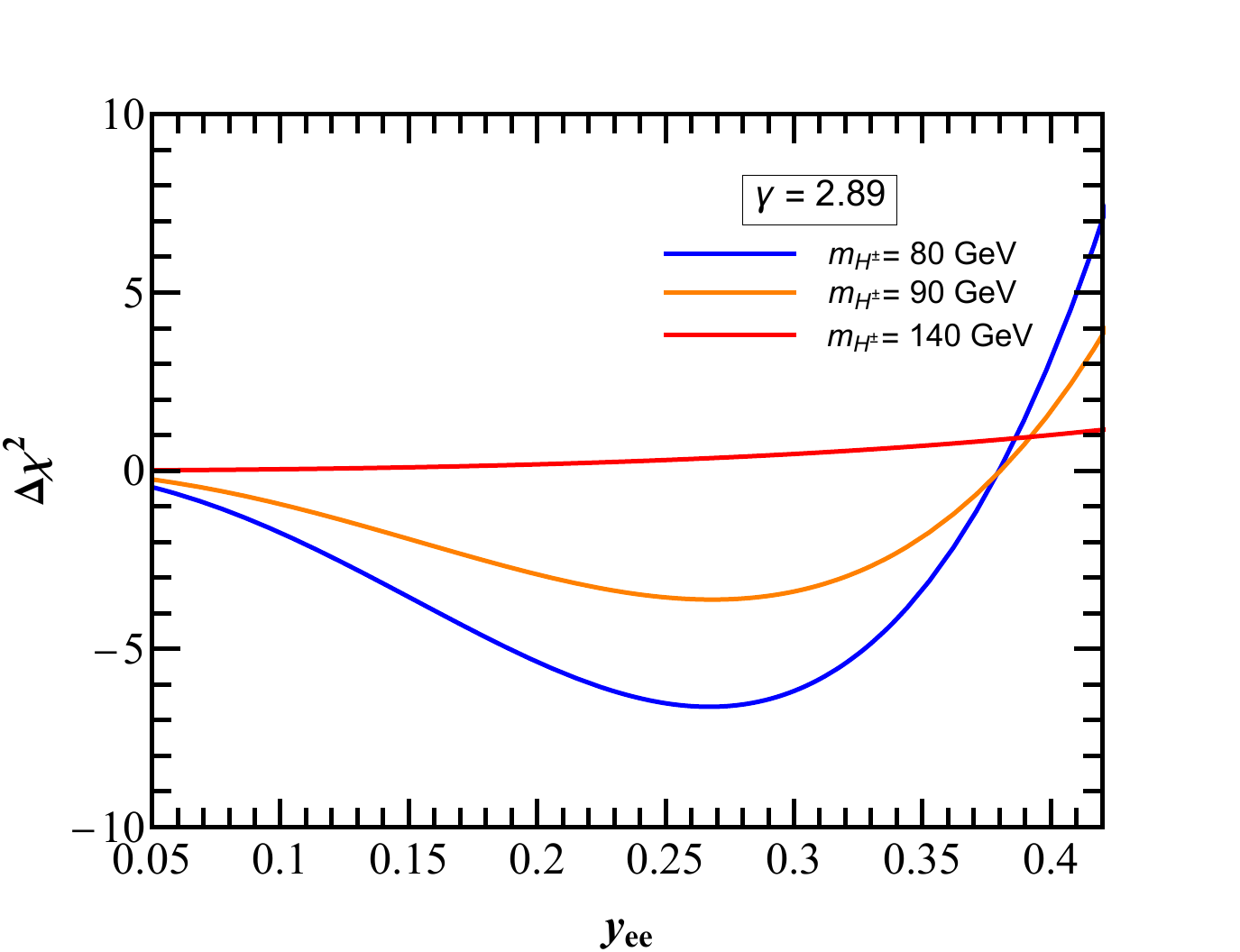} 
\includegraphics[scale=0.52]{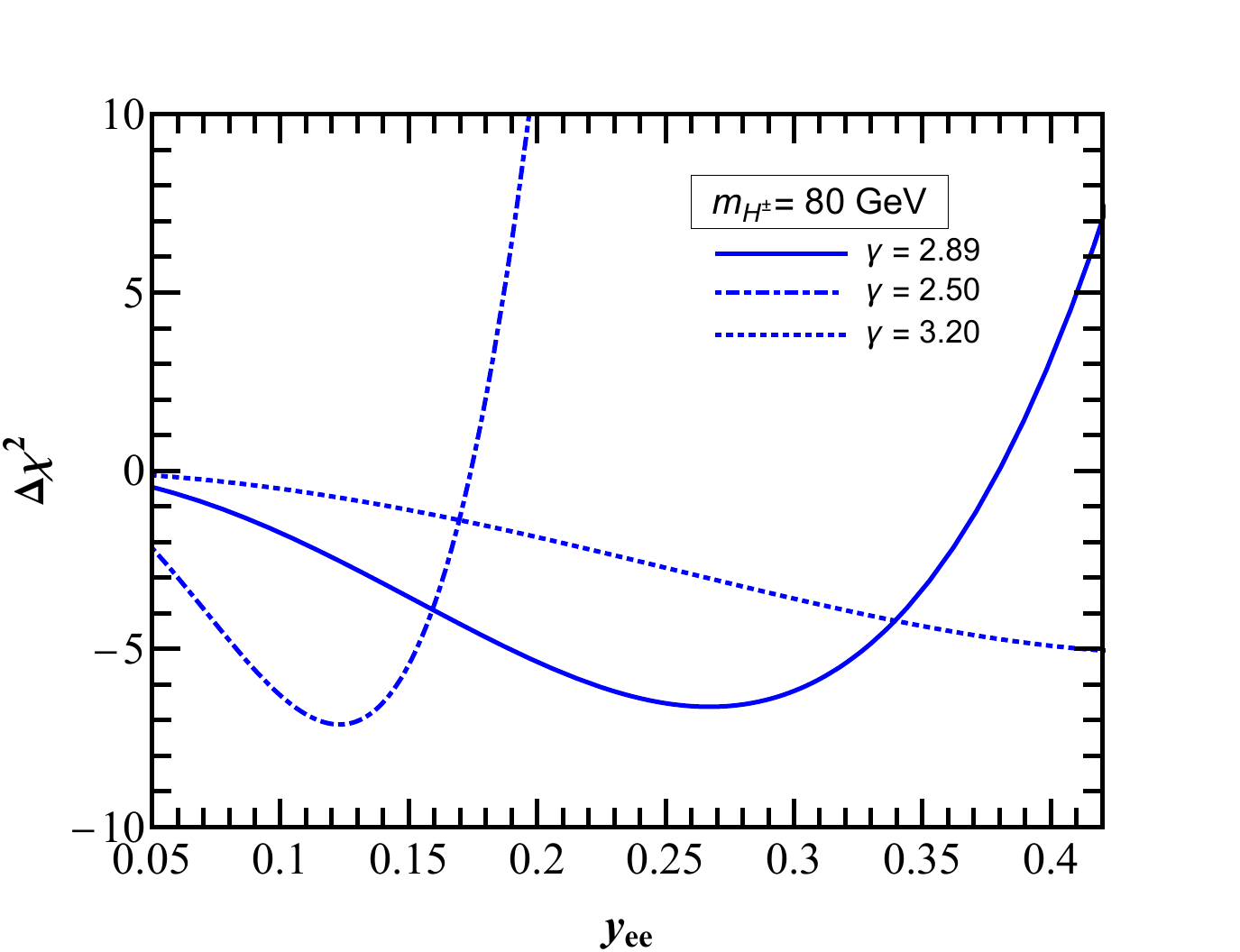} 
\caption{\footnotesize Variation of $\Delta \chi^2$ with $y_{ee}$ for different benchmark values of charged Higgs mass (left panel) and spectral index (right panel).}
\label{fig:ChiSq}
\end{figure}

Finally, we demonstrate our simulated results using the $ \Delta \chi^2 $ function. 
For our statistical analysis, effects of the presence of a charged Higgs resonance are estimated through the $\chi^2$ values.
We define the quantity as
\begin{equation}
\Delta \chi^2 = 100 \times \frac{\chi^2(\text{w/o}~ H^{\pm}) - \chi^2(\text{w} ~H^{\pm})}{\chi^2(\text{w/o} ~ H^{\pm})} \;,
\end{equation}
where $ \chi^2(\text{w/o}~ H^{\pm}) $ is the $\chi^2$ function that is defined in absence of charged Higgs resonance. 
In this function, the data set and relevant experimental error corresponding to  the astrophysical and atmospheric HESE are included, as described in Fig.~\ref{fig:Events}. Moreover, the contribution arising from the charged Higgs resonance is defined by $ \chi^2({\rm w}~ H^{\pm}) $ function. To quantify if the presence of the charged Higgs makes the fit to the current IceCube data better,
we compute $\Delta \chi^2$ values for different charged Higgs masses while varying the Yukawa $y_{ee}$, and this is presented in the left panel of Fig.~\ref{fig:ChiSq}. To qualitatively see the effect of astrophysical inputs, in the right panel of Fig.~\ref{fig:ChiSq} we show the $\Delta \chi^2$ in case of different characteristic spectral indices, namely $\gamma = 2.50, 2.89$, and 3.20, for a fixed charged Higgs mass of 80 GeV. From this plot it is clear that better fit to the current IceCube data can happen for a different type of neutrino flux spectrum, while other types do not necessarily modify anything. A relatively flatter spectrum with $\gamma= 2.50$ can have a similar or better fit to the IceCube results in presence of a charged Higgs resonance, albeit with lower Yukawa values. This happens since there is a greater probability of events in the 1-3 PeV bins with a flatter spectrum, even without the charged Higgs contribution. On the other hand, a spectrum sharper than the SM best-fit worsens the overall fit.

Our analysis revolves around the possible explanation of the super-PeV events observed at the IceCube through the charged Higgs resonance.
Notice here that the contributions arising from the charged Higgs resonance explain the observed events, hitherto absent in the astrophysical data, in the first or third super-PeV bins for $m_{H^{\pm}} = 80$, and $ 90$ GeV (see the orange and blue curves), respectively.  
This establishes the fact that an alternate better explanation of the IceCube PeV events can be made, considering the charged Higgs mass of 80 or 90 GeV with $y_{ee} \sim 0.28$ within the concerned model. Furthermore, we also show a case-study for the $\Delta \chi^2$ for $ m_{H^{\pm}} = 140$ GeV. One can see that as the latter fit is not good as compared to the remaining two cases, the corresponding charged Higgs mass is unable to provide any new insight.
After laying down the prescription of computing the IceCube events, we will now proceed to discuss the prospects of probing NSIs at the IceCube detector and in the process we will also demonstrate the constraints coming from various neutrino experiments.

\subsection{NSIs signature at IceCube}
\label{sec:nsi-IC}
Here, we discuss the testability of non-standard neutrino interactions arising from our model as has been pointed out in Sec.~\ref{sec:model}, with respect to the latest IceCube data~\cite{Schneider:2019ayi}.
%
%
\begin{figure}[!htbp]
\centering
\hspace{-6mm}
\includegraphics[scale=0.4]{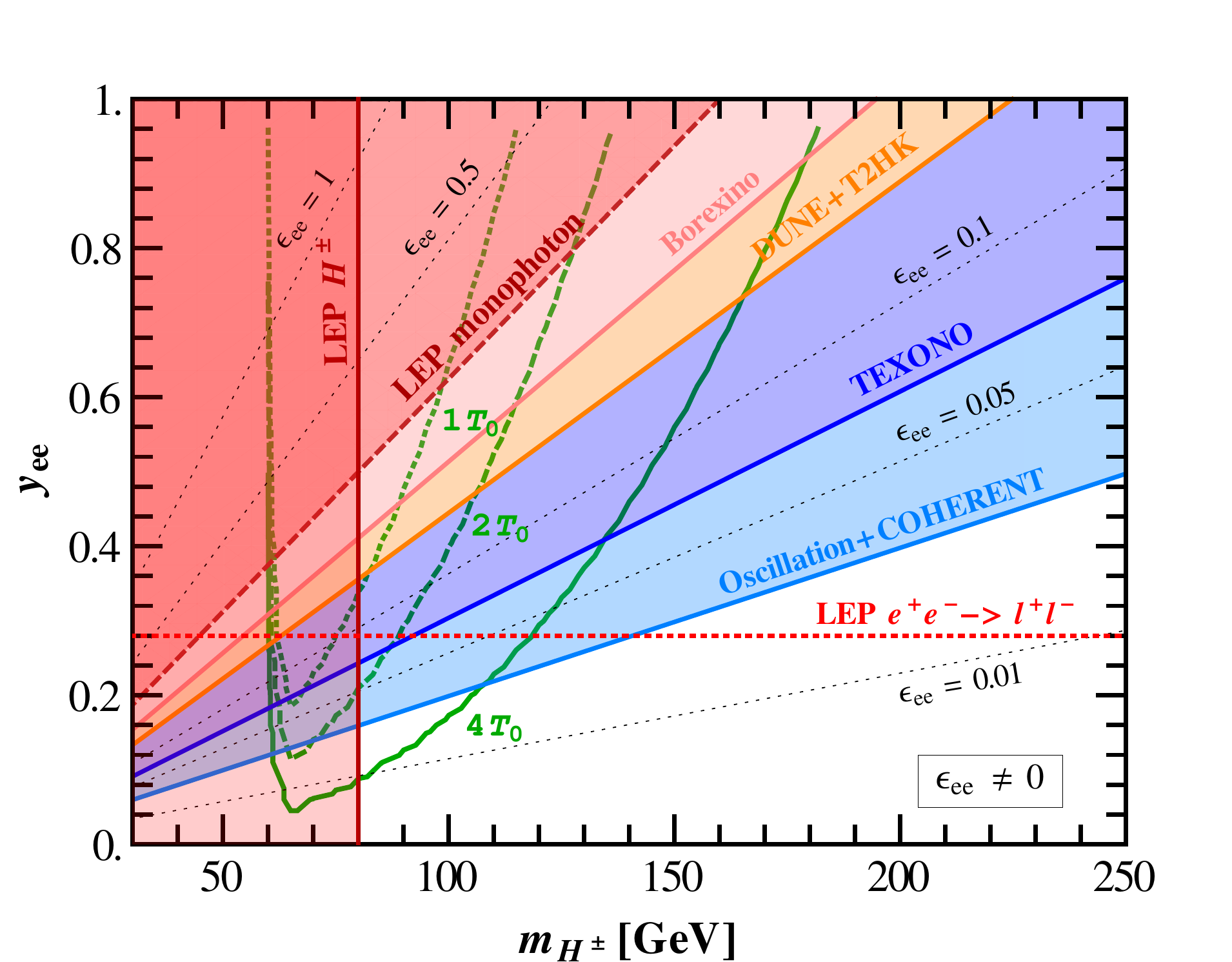}
\includegraphics[scale=0.41]{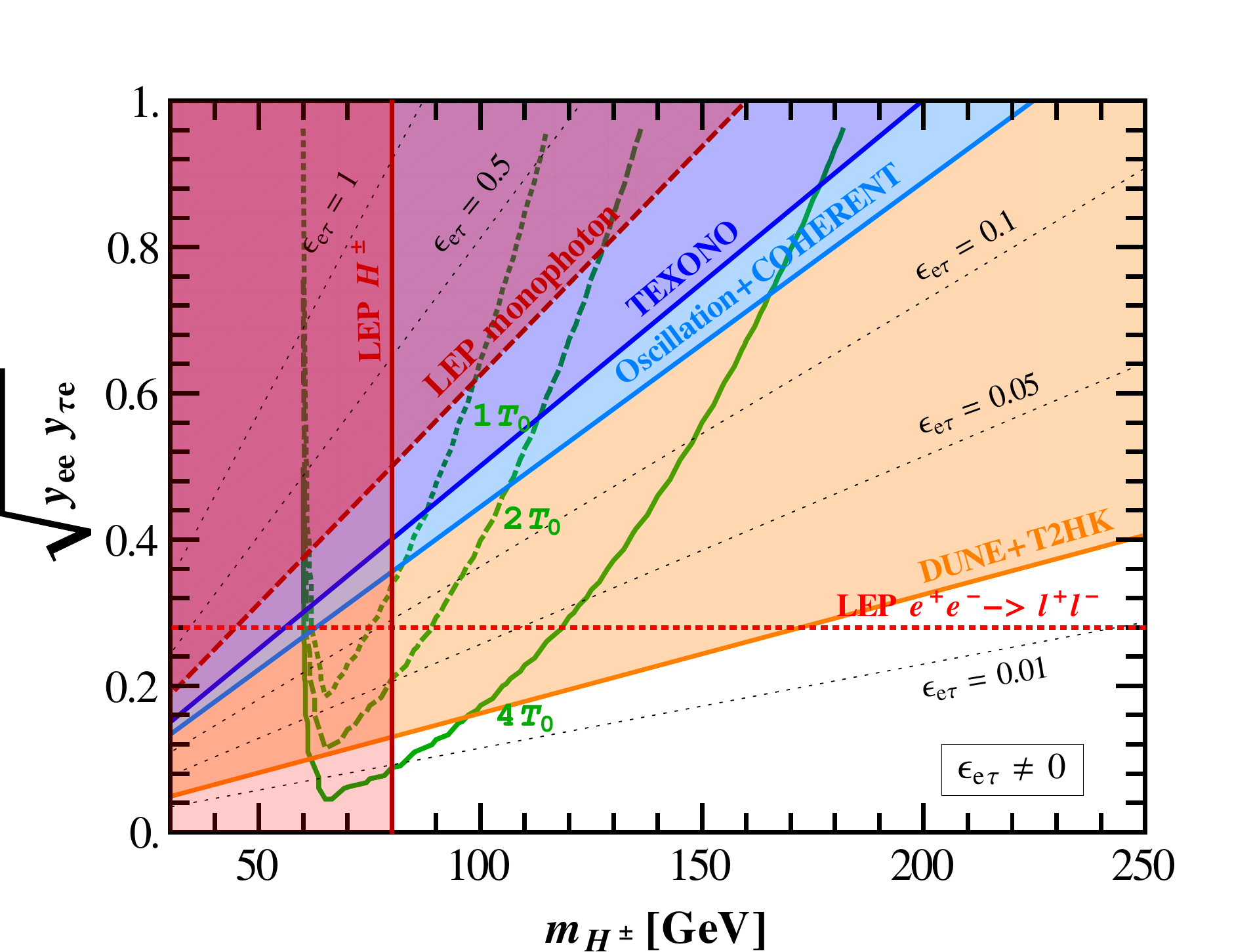} 
\caption{\footnotesize Expected sensitivity of IceCube  in the $(m_{H^{\pm}}, y_{ee})$ plane to test the NSI parameter $\epsilon_{ee}$ (see left panel), and ($(m_{H^{\pm}}, \sqrt{y_{ee} y_{\tau e}})$ plane to test the NSI parameter $ \epsilon_{e\tau} $ (see right panel),  considering total three neutrino events for the combined PeV energy bins~\cite{Schneider:2019ayi}, for different exposure times. Here time $T_0 ( = 2635$ days) represents the  current exposure time of the IceCube detector. 
The green dotted, dashed, and solid lines represent the iso-event contours set by the IceCube data for exposure times corresponding to $T_0, 2T_0, $ and $ 4T_0$, respectively.
Different benchmark values for  $ \epsilon_{ee} $ ($ \epsilon_{e\tau} $) are shown using the dotted black lines.
The constraints set by the LEP experiment \cite{Abbiendi:2013hk,LEP:2003aa,Fox:2011fx} are shown by the red colors for the charged scalar mass, from the measurement of  $ e^+ e^- \rightarrow  l^+ l^-$ cross-section, and monophoton signal, respectively.
The exclusion regions  coming from the neutrino experiments Borexino \cite{Agarwalla:2019smc}, combined DUNE+T2HK \cite{Acciarri:2015uup,Abe:2018uyc}, TEXONO \cite{Deniz:2010mp}, and combined Oscillation+COHERENT \cite{Coloma:2017ncl}  are presented at 90\% confidence level using the color code pink, orange, blue, and cyan, respectively (see text for more details). }
\label{fig:Contours} 
\end{figure}
%
%
In Fig.~\ref{fig:Contours}, we show the expected sensitivity of IceCube to test the NSI parameters $ \epsilon_{ee},  \epsilon_{e\tau} $ within our concerned framework.
In doing so, we consider total three observed neutrino events at the IceCube detector \cite{Schneider:2019ayi} in the energy range ($ 10^{6} \leq E_{\nu} \leq 10^{7}$ GeV) i.e.,  we have summed over the last nine energy bins of the IceCube event spectrum that lead to total three neutrino events (see Fig.~\ref{fig:Events} for details about the number of energy bins).
The IceCube sensitivity corresponding to `$ 1T_0 $' for the current exposure time $T_0 = 2635$ days has been shown using the dotted green line.
We also present two more curves by increasing exposure times of IceCube, i.e., for  $ 2T_0 $ and $ 4T_0 $ using the dashed and solid green lines, respectively~\footnote{Note that for $2T_0, $ and $ 4T_0$ we have only changed the IceCube exposure time as given by Eq.~\eqref{eq:ICevtNum}, while other configuration parameters of IceCube are kept the same.}.
Also, in order to have a more complete study, we include exclusion regions arising from various other neutrino experiments for both neutrino oscillation as well as neutrino scattering experiments at 90\% confidence level.
In the left panel,  $(m_{H^{\pm}} - y_{ee})$ parameter space has been presented which can test NSI parameter $ \epsilon_{ee} $ as given by Eq.~\eqref{eq:gennsimodel}. Similarly, we also show the parameter space in $(m_{H^{\pm}} - \sqrt{y_{ee}y_{\tau e}})$ plane to test NSI parameter $ \epsilon_{e\tau} $ following Eq.~\eqref{eq:gennsimodel}, in the right panel. 
Considering different benchmark values of NSI parameters $ \epsilon_{ee}, $ and $  \epsilon_{e\tau} $, various sensitivity lines have been shown with the help of Eq.~\eqref{eq:gennsimodel} using the dotted black lines.
Next we consider the limits set by the LEP experiment on charged Higgs mass and its coupling.
The LEP experiment searches for a charged Higgs through the process $e^+ e^- \to Z \to H^+ H^-$, where $H^{\pm}$ further decays to $\tau \nu$ sets a lower bound on the charged Higgs mass as $m_{H^{\pm}} > 80$ GeV~\cite{Abbiendi:2013hk}, as shown by the shaded red vertical band. 
Given the specific interactions in this model, this choice of $H^{\pm}\rightarrow \tau \nu$ LEP bound is quite conservative.
 This is because in our case the only decay mode that is possible for the charged Higgs is $H \rightarrow e \nu$, as all other charged Higgs couplings are suppressed by the tiny mixing of the two scalar doublets. Even the LEP constraint, $m_{H^\pm} > $ 80 GeV, or for that matter LHC constraints (based on hadronic or $\tau \nu$ decay, see~\cite{Akeroyd:2016ymd}) will not be applicable. Actually, all the constraints will be weaker in the absence of charged Higgs searches in $H \rightarrow e \nu$ mode.
The constraints coming from the measurement of $ e^+ e^- \to  l^+ l^-$ cross section  by the LEP experiment~\cite{LEP:2003aa} as well as from the dark matter searches in the monophoton signal $e^+ e^- \to \text{DM}~\text{DM}~\gamma$ \cite{Fox:2011fx} are shown by the dotted-red horizontal line and red-shaded regions (see Fig.~\ref{fig:Contours} for clarification), respectively. 
Note that for the process $ e^+ e^- \to \tau^+ e^-$, which in principle can occur by the mediation of neutral heavy scalars ($H$ and $A$), the cross section will be negligible in comparison to the process $ e^+ e^- \to  e^+ e^-$ simply because the $H(A)\tau e$ coupling is at least one order smaller than the $H(A)ee$ coupling. Therefore, inclusion of that will not affect the constrained parameter space. Moreover, the bound from LEP $e^+ e^- \to  l^+ l^-$ will not have any explicit dependence on $H^{+}$, at tree level, due to absence of the corresponding couplings. However, it does have an explicit dependence on the neutral scalars, but the considerations of constraints coming from electroweak precision observables, namely $S$, $T$ parameters relate these masses. 
We work with the highest neutral boson masses possible from the electroweak precision observables (see discussions in~\cite{Dey:2018yht, Machado:2015sha}) such that there remains a breathing space for Yukawa $y_{ee}$ from the LEP bound. Since we are considering quite a conservative value for $m_{H^{\pm}}$ (which results in fixed masses of other neutral BSM scalars, $m_{H,A}$) we show it just as a dotted horizontal red line in Figs.~\ref{fig:Contours}. A relatively loose constraint on $y_{ee}$ helps generating sizeable NSI parameter values while satisfying the LEP $e^+ e^- \to  l^+ l^-$ bounds.
The exclusion regions coming from  the solar neutrino experiment Borexino~\cite{Agarwalla:2019smc} is presented using the pink shaded region. Notice that the Borexino collaboration~\cite{Agarwalla:2019smc} restricts their analysis for the flavour-diagonal neutral current interactions using neutrino-electron elastic scattering process (i.e., for $ \nu_{e} e, \nu_{\tau} e $ couplings). Therefore, they only set bounds for the diagonal NSIs $ \epsilon_{ee},$ and $ \epsilon_{\tau\tau} $. The corresponding results for  $ \epsilon_{ee} $ is shown in the left panel. 
On the other hand, the neutrino-electron scattering experiment TEXONO~\cite{Deniz:2010mp} sets bound on all the  NSI parameters involving electrons. The bounds arising from TEXONO collaboration~\cite{Deniz:2010mp} for both $ \epsilon_{ee},$ and $ \epsilon_{e\tau}  $ are shown using the blue shaded regions. Besides this, the sensitivity arising from the combined analysis of neutrino oscillation experiments together with the  recent observation of coherent neutrino–nucleus
scattering process by the COHERENT collaboration is shown by the cyan shaded region. We show the exclusion region for $ \epsilon_{ee},$ and $ \epsilon_{e\tau}  $ adopting the bounds discussed in~\cite{Coloma:2017ncl}.
Moreover, the bounds coming from the combined analysis of the next-generation neutrino oscillation experiments DUNE+T2HK \cite{Acciarri:2015uup,Abe:2018uyc} is shown by the shaded orange regions.
To find the expected sensitivity of both  NSI parameters in case of DUNE+T2HK, we perform the combined numerical simulation of both the experiments using the \texttt{GLoBES} packages~\cite{Huber:2004ka,Huber:2007ji} along with the required auxiliary files presented in Ref.~\cite{Alion:2016uaj} for DUNE and the detector simulation files in Ref.~\cite{Liao:2016orc} for T2HK. 
We use the \texttt{GLoBES} extension file \texttt{snu.c} as has been presented in Refs.~\cite{Kopp:2006wp,Kopp:2007ne} to incorporate NSIs. On the other hand,  numerical values for the standard three flavour neutrino oscillation parameters have been adopted from the latest global-fit study~\cite{deSalas:2020pgw} for the normal neutrino mass hierarchy. In this combined analysis, sensitivity of $ \epsilon_{ee}$ is found in the range ($ -0.15, 0.15 $), whereas sensitivity of $ \epsilon_{e\tau}$ is noticed $ \leq 0.02 $ at 90\% confidence level. 
Now, with the help of Eq.~\eqref{eq:gennsimodel}, we transform these bounds, considering $ \epsilon_{ee} = 0.15$ and  $ \epsilon_{e\tau} = 0.02$, in the  charged Higgs mass vs coupling planes as shown in Fig.~\ref{fig:Contours}.

From both the panels of Fig.~\ref{fig:Contours}, we notice that the tight LEP bounds on charged scalar mass rule out parameter space $ m_H^{\pm} <  80$ GeV as shown by the shaded red regions. Furthermore, the LEP constraint of $ e^+ e^- \rightarrow  l^+ l^-$ rules out $ y_{ee} > 0.28$, which we show using the dotted red line.
Investigating the left panel of Fig.~\ref{fig:Contours}, it can be observed that the constraint coming from the current IceCube data $ 1T_0 $  on $ \epsilon_{ee} $ (see dotted green curve) can be ruled out in comparison to LEP bounds on charged scalar mass as well as by noticing  bounds from the Oscillation+COHERENT analysis. 
This is due to the fact that for NSI parameter $ \epsilon_{ee} $, Oscillation+COHERENT provides the most stringent bounds as compared to other neutrino experiments as shown by the cyan coloured region. 
The same findings remain true for $ 2T_0 $ IceCube data as shown by the dashed green curve. However, the future IceCube data with $ 4T_0 $ exposure time will be able to explore a parameter space for the charged scalar masses between 80 GeV to 105 GeV and coupling as small as $ \sim 0.1 $. This further shows that in future IceCube will be able to test NSI parameter $ \epsilon_{ee} \sim 0.01 $. 
From the right panel, one sees that the sensitivity arising from both $ 1T_0 $, and $ 2T_0 $ IceCube data will be ruled out by the combined sensitivity coming from the DUNE+T2HK analysis. Nonetheless, an exposure time of $ 4T_0 $ IceCube shows the best sensitivity for the charged Higgs mass around 90 GeV and for a coupling around 0.1 (see the solid green line).  
It is important to note here that the future IceCube data with $ 4T_0 $ exposure time will have the capability to surpass all the bounds appearing from different neutrino experiments for the given charged scalar model.
We end this section by drawing a parallel between our model and more elaborate 2HDMs. From the NSI perspective, more elaborate 2HDM scenarios can potentially generate sizeable NSIs. However, significant NSIs are only possible when charged Higgs mass is around 100 GeV, which is not favourable from multiple constraints in elaborate 2HDM models, if quark couplings are present.
In our study we maintained relevant constraints from various sectors while performing the IceCube analysis. Even from the IceCube point of view, the charged Higgs produced by the ultra-high energy incoming neutrinos can decay to jets in elaborate 2HDMs due to quark couplings of the BSM scalars. In that case dominant hadronic shower from the charged Higgs can, in principle, shadow the Glashow resonance peak at 6.3 PeV. Some clever methods are needed to be devised to distinguish the two. Surely, in a leptophilic scenario this effect will not be so prominent. On the other hand for IceCube to observe any effect of charged Higgs bosons forming a resonance before 10 PeV, the charged Higgs mass should be around 100 GeV. Most of the elaborate 2HDMs cannot arrange for such a lighter charged Higgs keeping into account all other existing constraints. As an example we can show that in the type-II and type-Y 2HDM the flavour constraints push the charged Higgs mass beyond 600 GeV~\cite{Chowdhury:2017aav}. In that respect our set-up is insulated from these issues owing to the very specific leptophilic charged Higgs.

\section{Conclusion}
\label{sec:concl}
The IceCube collaboration has reported a number of neutrino events in the PeV energy range, one of which is even around the Glashow resonance energy.
In the absence of a full-proof model of astrophysical neutrino flux and low statistics in the IceCube data the exact explanation of these events are not possible. This work is a humble attempt to address this observation in the context of a BSM framework. Here, a neutrinophilic 2HDMs ($\nu$2HDM) has been adopted, where an extra Higgs doublet has been added in the SM. The charged Higgs, present in the model, is responsible for a Glashow-like resonance, furnishing a plausible explanation of PeV event(s).
On top of this the scenario also leads to neutrino non-standard interactions (NSIs), again thanks to the relevant couplings between the neutrinos and the charged Higgs. 
Clearly, numerous constraints arising from different particle physics experiments has to be satisfied. Considering various limits, we find that this model allows charged scalar mass $\sim \mathcal{O}(100)$ GeV that provides only two sizeable NSIs $ \epsilon_{ee} $, and $ \epsilon_{e \tau} $.
In this charged scalar mass limits, we adopt two benchmark values 80, 90 GeV which give almost an order of magnitude enhancement in the measurements of cross-section compared to the SM cross-section, as shown in Fig.~\ref{fig:CrossSection}. Besides this, these benchmark values also explain excess PeV neutrino events as summarizes in Fig.~\ref{fig:Events}.
We further explore the sensitivity of the current IceCube data to constrain the parameter space in the charged scalar mass vs. coupling constant plane. It has been observed that other neutrino experiments set more stringent limits compared to the latest bounds arising from the IceCube data. 
We summarize these notable results in Fig.~\ref{fig:Contours}. Indeed, with an increase in exposure time of the IceCube detector, i.e., with four times of the existing exposure, in future IceCube can surpass all the bounds set by the other neutrino experiments and can also test NSIs as small as $ \sim 0.01 $.
The set-up prescribed here can pave the way for further explorations of plethora of BSM scenarios with charged scalars and its requisite interactions in the IceCube as well as lower energy neutrino experiments. From the observational point of view although we resort to larger exposure times at IceCube detector, next generation upgrades, like IceCube-Gen2~\cite{Aartsen:2020fgd} owing to its larger effective volume of 8 km$^{3}$ will require much less exposure times to probe such scenarios. Parallelly, future water-based detectors like KM3NeT~\cite{Adrian-Martinez:2016fdl}, Baikal-GVD~\cite{Avrorin:2018ijk} etc. can also play a pivotal role in search of these scenarios.   

\acknowledgments
UKD acknowledges the support from Department of Science and Technology (DST), Government of India under the grant reference no. SRG/2020/000283.
NN is supported by the postdoctoral fellowship program DGAPA-UNAM, CONACYT CB-2017-2018/A1-S-13051 (M\'{e}xico) and DGAPA-PAPIIT IN107118.
SS thanks UGC for the DS Kothari postdoctoral fellowship grant with award letter No.F.4-2/2006 (BSR)/PH/17-18/0126.

\bibliographystyle{JHEP}
\bibliography{refIC.bib}

\end{document}